\documentclass{appolb}
\usepackage{lineno}
\usepackage{subfigure}
\usepackage{adjustbox}
\usepackage{amsmath}  % needed for \tfrac, \bmatrix, etc.
\usepackage{amsfonts} % needed for bold Greek, Fraktur, and blackboard bold
\usepackage{graphicx} % needed for figures
\usepackage{multirow}
\usepackage{gensymb}
\usepackage{textcomp}
\usepackage{xcolor}
\usepackage{subfigure}
\usepackage{setspace}
\usepackage[utf8]{inputenc}

\RequirePackage[colorlinks,citecolor=blue,urlcolor=blue,linkcolor=blue]{hyperref}
\RequirePackage{url}

\begin{document}
\title{Performance study of a Time of Flight Method  used for cosmic ray detection}
%\thanks{Presented at ...}%
\author{AL\.{I} YILMAZ
\address{Department of Electrical and Electronics Engineering, Giresun University, 28200, Giresun, Turkey.}
\\
}
\maketitle
\begin{abstract}
Time of Flight methods have been rapidly developed and used in many experiments recently for determination of particle direction, identification of particles and energy resolutions.
This paper describes a method of time-mark determination on the reconstruction algorithm, based on the sampled signal, used for time-of-flight measurements. This method was developed for distinguishing the signals by fitting to pulse shape which were received from scintillator detector with a silicon photomultiplier readout have been developed for a cosmic ray counter telescope. The method was verified using experimental data taken in the location $40\degree54'52''N$ and $38\degree19'26''E$ with the elevation of 30 m above the sea level. The data samples were acquired by the counters which have a scintillator with dimensions of 20$\times$20$\times1.4~cm^3$, optically coupled from one side to silicon photomultiplier, then the signals readout by fast sampling digitizer board Domino Ring Sampler Board version 4. The method can reconstruct each pulse even for multiple events without losing the count within the small time window. Using this method 4.969 ns time-of-flight value were established and the rise times for scintillation counters, named Tile 1 and Tile 2, were measured about $6.27 \pm 0.16~ns$ and $4.979\pm0.165~ns$, respectively.
\end{abstract}

%\PACS{PACS numbers come here}
  
\section{Introduction} \label{sec:intro}
Precisely measuring the timing information of a particle in an experiment would allow one to successfully reconstruct physical events. Time of Flight (TOF) methods have been developed and used in many fields such as high energy physics experiments~\cite{BARWICK199734,ALICI2014288}, astroparticle detectors~\cite{RATCLIFF20081,IORI2014265}, TOF cameras~\cite{MUFTI2011720, GILLES2015S823}, TOF-PET detectors~\cite{RONZHIN2013109, JORAM2013586} for identifying the particle species, determining the direction of  the particles and energy measurements.  
The counter telescope is useful in researches which is a relatively a few number of desired particles has to be counted in a major unwanted  background. The analyzing the coincidence time, passing through the counters telescopes, is practical to remove the spurious counts registered because of the accidental coincidence of counts in the each detectors.
While comparing the Silicon PhotoMultipliers (SiPMs) with traditional PhotoMultiplier Tubes (PMTs), which have been widely used  in the past experiments for decades, SiPMs have some advantages such as advanced photon detection efficiency, compactness, excellent single photon time resolution, insensitive to magnetic fields and inexpensive price. Due to these advantages, they have been started to use in a wide range of physics experiments nowadays. A SiPM signal, is constituted of a few different avalanche events formed at arbitrary times.
The difficulties in achieving high accuracy on the time-marking of the signal and TOF resolutions of a counter telescope with existing algorithms consist the subsequent subjects: \\
%\begin{itemize}
%	\item
	 \textit{Variation of the baseline:} If the dark count rate increases considerably then peaks of the signal are mostly overlying on each other within a time latency. Therefore, the method requires to detect and evaluate signal starting time for different heights while working at higher over-voltages. \\
%	\item 
	\textit{Electrical noise:} The signal to noise ratio needs to be at lowest point 5 in the worst conditions (minimum over-voltages and higher electronic noise). If the primary signals will be evaluated mostly which yields an ambiguity of 20\% on the signal size while computing directly thus a fitting algorithm is required.\\
%	\item 
	\textit{Very different signal heights and widths:} Due to the proposed method can be able to adjust itself while measuring various properties without a user concerning to change and optimize the parameters for any evaluation, very different signal heights and widths vary strongly with over-voltage and SiPM type.~\cite{Putignano2012}.
%\end{itemize}

The rest of the paper is organized as follows. In the next section is described used experimental setup for signal reconstruction. In Section 3, a method developed to extract time information of the signal coming from scintillation counters without varying difficulties listed above. Finally, an application of this method to the experimental data is constructed and the results of its is given.

\section{Experimental Details}

\subsection{Description of the setup}
\label{subsec:DAQ}

\begin{figure}[htb]
\centerline{%
\includegraphics[width=1\textwidth]{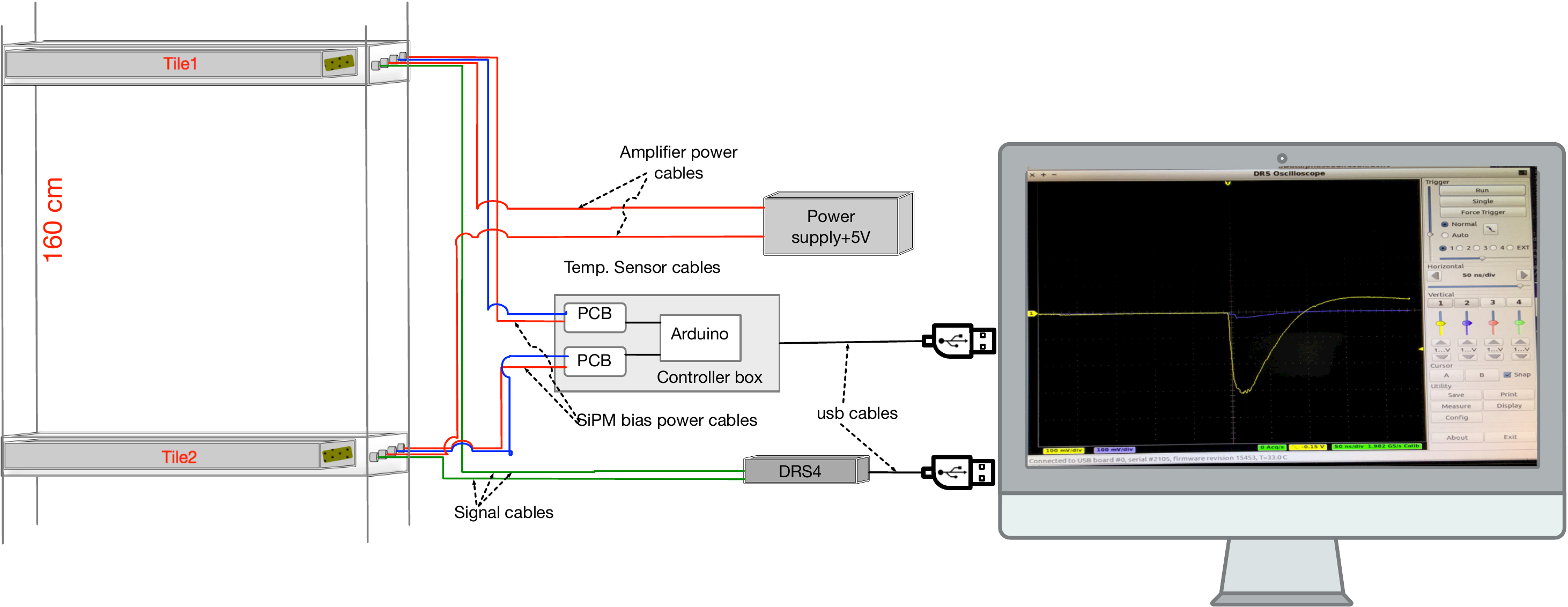}}
\caption{Schematic view of data acquisition.}
\label{fig:detectorSchematic}
\end{figure}

 The hardware of the scintillation counter telescope  constructed with two identical scintillator plates, called $Tile~1$ and $Tile~2$, separated by $160~cm$ apart as seen in Figure~\ref{fig:detectorSchematic}. Each counter box consist of a KURARAY organic scintillator plate ($20\times20\times1.4~cm^3$). The scintillator has excellent properties, in the view of getting the accurate time information, like yielding the light in blue region of the spectrum, the emission peak is around 430 nm~\cite{kuraray}. Each scintillator panel wrapped in Tyvek paper for diffusing the reflection, and one SensL SiPM ($3\times3$ mm$^2$) contacted to read the produced signal. This SiPM is able to produce a sharp output pulse of $<2~ns$ at Full Width at Half Maximum (FWHM). The bias voltage of this device is about $27.5~V$ and the dynamic range over the breakdown voltage is about $2~V$. Hence the filling factor of the device is 64\%, the gain is $2.3\times10^6$~\cite{sensl}. The produced signal is digitized by Domino Ring Sampler Board (DRS4), developed by Stefan Ritt~\cite{RITT2004470}. 
 The Data Acquisition (DAQ) program is based on the setup schematically shown in Figure~\ref{fig:detectorSchematic}. It is managed by a shell script that controls the two main C\texttt{++} programs. One of the program is controlling the Arduino so that reads temperatures from the SiPM readout circuit and adjust the operating voltage in order to keep constant the gain of the SiPM on each tile. The other program manages the Domino Ring Sampler board $v4$ (DRS4) which digitizes the signal detected by the SiPM and stores in ROOT binary format for further analysis.
 The DAQ is based on waveform sampling at 2 GS/s, covering a $2.5~\mu s$ window. This detector can be used to select horizontal tracks for detecting tau shower produced by the neutrino interacting in Earth crust~\cite{iori2004test, IORI2008151, Yilmaz:2017gub}.  So the TOF resolution of the selected tracks can be achieved to $0.5~ns$.

\subsection{Description of the reconstruction method}

The reconstruction program, is based on ROOT analysis package~\cite{brun1997root} using  C\texttt{++} programming language, represents a necessary tool for the analysis of the data. 
The method optimized for high sensitivity while determining the peak level  and rise time of the signal are required to effectively use the peak height readout option. The method explained in this section is developed to work on a continuous SiPM signal.

\subsubsection{Least Square Method}\label{subsec:leastSquare}

A function was written in the analysis program in order to determine the the starting point of the signal. This point gives the timing information of the particle so that a mathematical procedure can be applied for estimating the best-fitting curve to a dedicated set of points by minimizing the sum of the squares of the offsets (”the residuals”) from the curve. By substituting a set of $N$ data points ($x_{i}, y_{i}$) to a straight-line function

\begin{eqnarray}
	y(x) = mx + c 
\end{eqnarray}
where $y(x)$ and $x$ are holding the voltage and time informations of the SiPM signal, respectively.
The equation is generally called \textit{least square method} or \textit{linear regression}~\cite{Margulies1968}. Assuming that the uncertainty $\sigma_{i}$ associated with each measurement $y_{i}$ is known, and that the $x_{i}$'s
(values of the dependent variable) are known exactly. The chi-square, $\chi^{2}$, function is used to compute how well the model matches up with the data.

\begin{eqnarray}
	\chi^{2}(m,c) & = & \sum_{i = 0}^{N - 1} \Big[\frac{y_{i} - mx_{i} - c}{\sigma_{i}}\Big]^{2}
	\label{eqn:meritFunction}
\end{eqnarray}
When the errors of a measurement are spread out normally, then this function will yield maximum likelihood coefficient evaluations of $m$ and $c$; if they are not spread out normally, then the evaluations are not maximum likelihood but might be advantageous to use in a practical sense.
 In order to determine $m$ and $c$ the Eq.\ref{eqn:meritFunction} is minimized. At its minimum, derivatives of $\chi^{2}(m, c)$ with respect to $m, c$ become zero:
 \begin{eqnarray}
 	\frac{\partial \chi^{2}}{\partial c} & = & -2 \cdot \sum_{i = 0}^{N - 1} \frac{y_{i} - mx_{i} - c}{\sigma_{i}^{2}} = 0 \nonumber\\
	\frac{\partial \chi^{2}}{\partial m} & = & -2 \cdot \sum_{i = 0}^{N - 1} \frac{x_{i}\cdot(y_{i} - mx_{i} - c)}{\sigma_{i}^{2}} = 0 
	\label{eqn:meritFunction2}
 \end{eqnarray}
These conditions can be revised in a suitable form given in the following sums:
\begin{eqnarray}
	\xi_{\sigma} \cong \sum_{i = 0}^{N - 1} \frac{1}{\sigma_{i}^{2}} \qquad \xi_{x} \cong \sum_{i = 0}^{N - 1} \frac{x_{i}}{\sigma_{i}^{2}} \qquad \xi_{y} \cong \sum_{i = 0}^{N - 1} \frac{y_{i}}{\sigma_{i}^{2}} \nonumber\\
	\xi_{xx} \cong \sum_{i = 0}^{N - 1} \frac{x_{i}^{2}}{\sigma_{i}^{2}}  \qquad \xi_{xy} \cong \sum_{i = 0}^{N - 1} \frac{x_{i}y_{i}}{\sigma_{i}^{2}} 	
	\label{eqn:meritFunction3}
\end{eqnarray}
the Eq.\ref{eqn:meritFunction2} can be rewritten by using the definitions in the Eqs.~\ref{eqn:meritFunction3}; 
\begin{eqnarray}
	 m\xi_{x} + c\xi_{\sigma} & = & \xi_{y} \nonumber \\
	m\xi_{xx} + c\xi_{x}  & = & \xi_{xy} 
	\label{eqn:meritFunction4}
\end{eqnarray}
The result of these two equations in two unknowns is computed as
\begin{eqnarray}
	\triangle & \cong & \xi_{\sigma}\xi_{xx} - (\xi_{x})^{2} \nonumber \\
	c & = & \frac{\xi_{xx}\xi_{y} - \xi_{x}\xi_{xy}}{\triangle}, \quad
	m = \frac{\xi_{\sigma}\xi_{xy} - \xi_{x}\xi_{y}}{\triangle}
	\label{eqn:meritFunction5}
\end{eqnarray}
The solution for the best-fit model parameters $m$ and $c$ is given in the Eq.\ref{eqn:meritFunction5}. The expected uncertainties in the calculation of $m$ and $c$ should also be evaluated for the reason that the measurement errors in the data naturally bring some of uncertainty in the setting of those parameters. Assuming the data set are independent, then each value puts its own bit of uncertainty to the parameters. Taking into account the spread of errors indicates that the variance $\sigma_{f}^{2}$ in the value of any function is as follow:
\begin{eqnarray}
	\sigma_{f}^{2} & = & \sum_{i = 0}^{N - 1} \sigma_{i}^{2} \cdot \Big(\frac{\partial f}{\partial y_{i}}\Big)^{2}
	\label{eqn:meritFunction6}
\end{eqnarray}
The derivatives of $m$ and $c$ in respect of $y_{i}$ might be directly calculated from the solution for the straight line equation:
\begin{eqnarray}
	\frac{\partial c}{\partial y_{i}}  =  \frac{\xi_{xx} - \xi_{x}\xi_{x_{i}}}{\sigma_{i}^{2}\cdot \triangle}, \quad \frac{\partial b}{\partial y_{i}}  =  \frac{\xi_{\sigma}\xi_{x} - \xi_{x_{i}}}{\sigma_{i}^{2}\cdot \triangle}
	\label{eqn:meritFunction7}
\end{eqnarray}
after the summation over the points as in the Eq.\ref{eqn:meritFunction6};
\begin{eqnarray}
	\sigma_{c}^{2}  =  \frac{\xi_{xx}}{\triangle}, \quad \sigma_{m}^{2}  =  \frac{\xi_{xx}}{\triangle} 
\end{eqnarray}
which are called the variances in the estimates of $m$ and $c$, respectively~\cite{William2007, weisstein2002least}.

\section{Results} \label{sec:results}
\subsection{Application of the Time of Flight Method to the experimental data}
\label{subsec:TOFmethod}
\begin{figure}[htb]
\centerline{%
\includegraphics[width=0.7\textwidth]{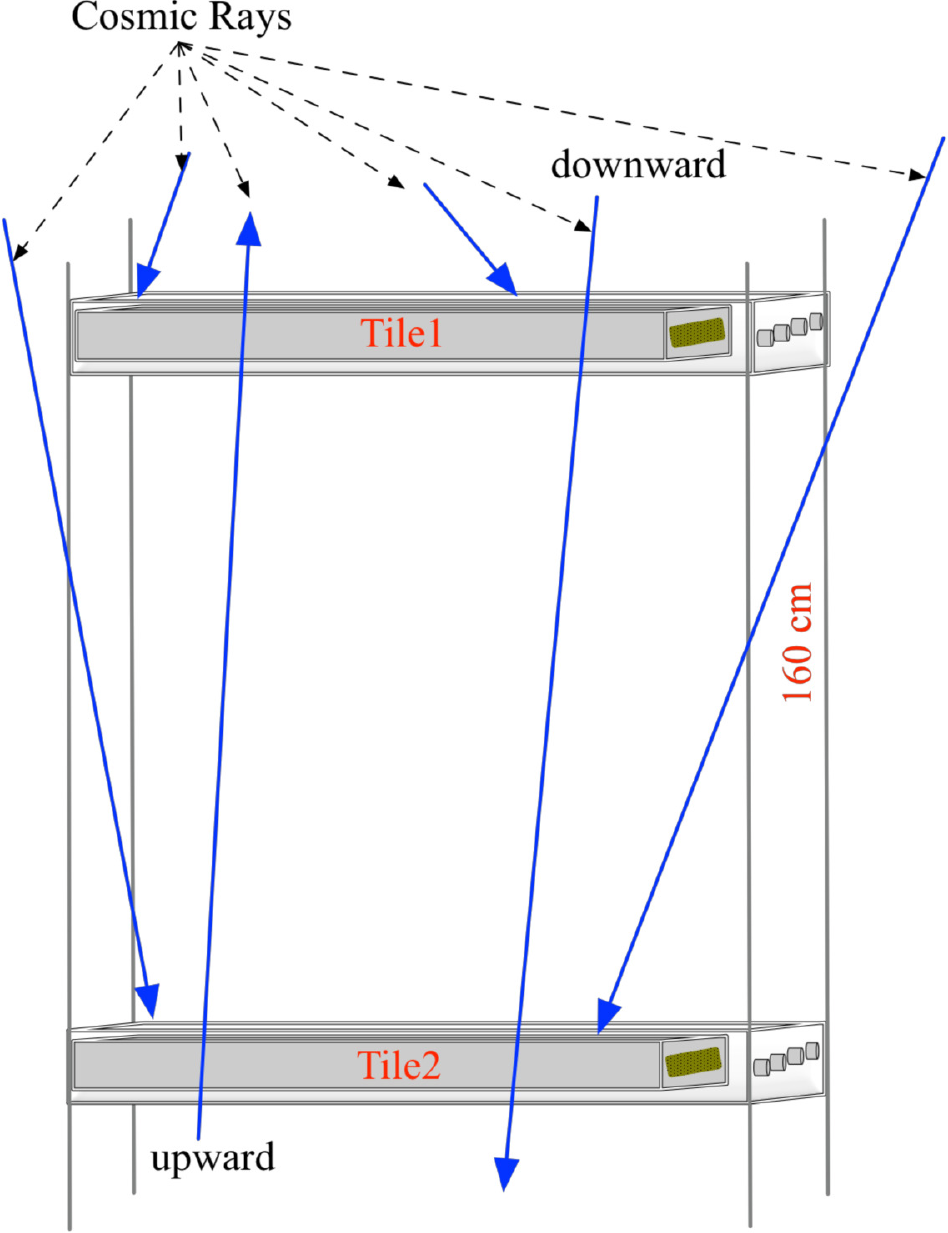}}
\caption{Schematic view of a counter telescope.}
\label{fig:tof}
\end{figure}

The counter telescope is useful in researches where a relatively a few number of desired particles has to be counted in a major unwanted  background. The analyzing the coincidence time is efficient way of removing the spurious counts occurred due to the accidental coincidence in the each detectors.
 TOF is a method measuring the time that it takes for a particle to travel a distance of a medium. While a particle hitting one of the scintillator tile as shown in Figure~\ref{fig:tof} that starts the time counter and it will be stopped when the particle hits with the second scintillator tile. 

\begin{figure}[hbt] 
 \centering
    \includegraphics[width=1\textwidth]{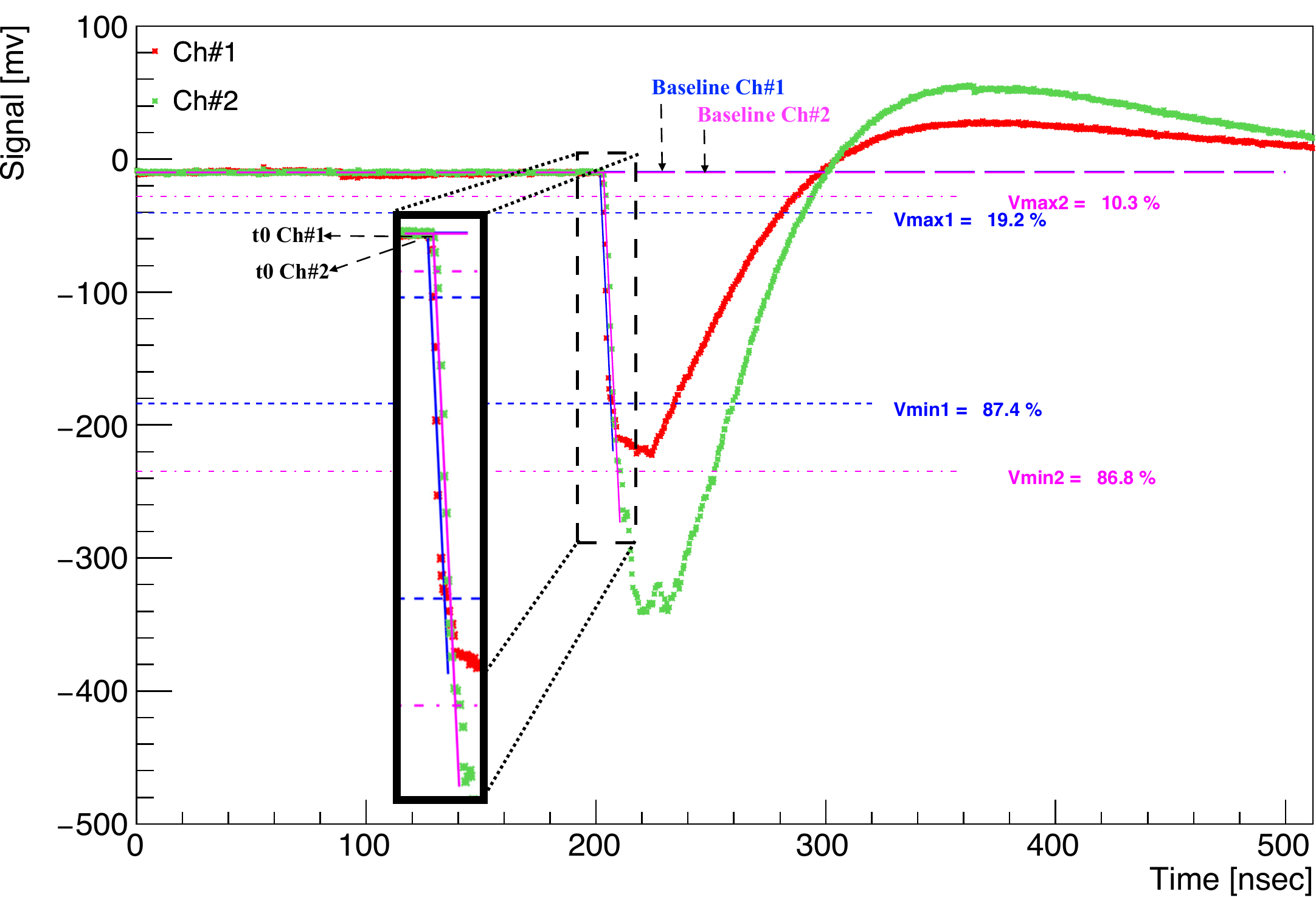}
  \caption{Description of proposed TOF calculation method used in this study. Tile separation is 160 cm. Magenta and blue lines indicate the fitted lines according to the method works forward direction from 10\% until 90\% of the amplitude.}
\label{fig:tof_method}
\end{figure}

TOF method is also useful to discriminate the particle direction (\emph{upward or downward}). Firstly, if the particle hit the Tile 1, then hits to Tile 2 means the particle is going to \emph{downward} direction, the expected result of the time  difference (\emph{$t_{0}ch2-t_{0}ch1$}) should  be \emph{positive}, if not, the particle is going to \emph{upward} direction.
\\ Figure~\ref{fig:tof_method} shows the algorithm used for calculating the time of flight method which have mentioned in section section~\ref{subsec:leastSquare}. The signals are registered by using DRS4 board and and then plotted by \emph{ROOT} program to observe the time difference between two tiles. If the signal is greater than the threshold voltage ($\sim40~mV$), then the program estimates the baseline of each signal in order to be insensitive to the baseline instability. After that, if the signal is inside the gate ($180~ns - 250~ns$), the analysis program firstly finds the peak level (absolute minimum) of the signal is the 100\% of the amplitude, then goes backward until \textit{safePoint} (which is at least $3\sigma$ below the baseline). Next, the program stores those points for being used in fit method. The program marks the 10\% and 90\% percents of the amplitude, which are not fix number of points, is just below the threshold value and estimated safePoint. The program fits a line equation using these registered points using the method explained in section~\ref{subsec:leastSquare}, and  crosses its own baseline axis (time axis). This point registered as \emph{$t_{0}$}. The difference of \emph{$t_{0}$} gives us the time-of-flight between two counters. 

\begin{figure}[!hbt] 
 \centering
    \includegraphics[width=1\textwidth]{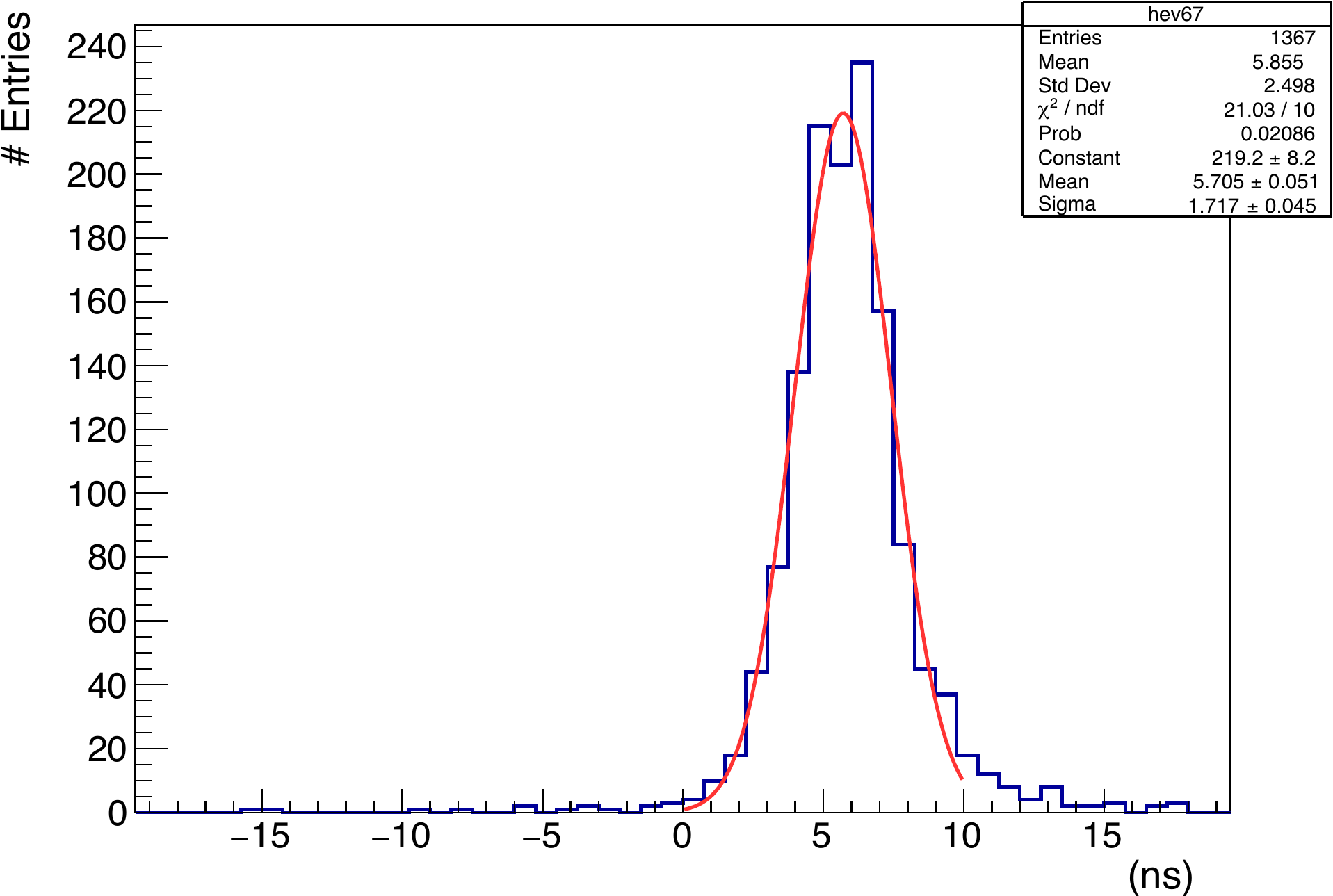}
  \caption{Time of Flight distribution of the counter telescope for tile separation is $160~cm$. The peak around $+5~ns$ is due to the particle coming from $Tile~1$ to $Tile~2$ (downward direction).}
\label{fig:TOF21}
\end{figure}

Figure \ref{fig:tof_method} shows the scheme of this algorithm. The computed TOF is about $5 \pm 1.7~ns$ for a $downward$ going particle is shown in Figure~\ref{fig:TOF21}.  

\subsection{Rise Time Test} \label{subsec:RiseTime}
\begin{figure}[hbt] 
 \centering
    \includegraphics[width=1\textwidth]{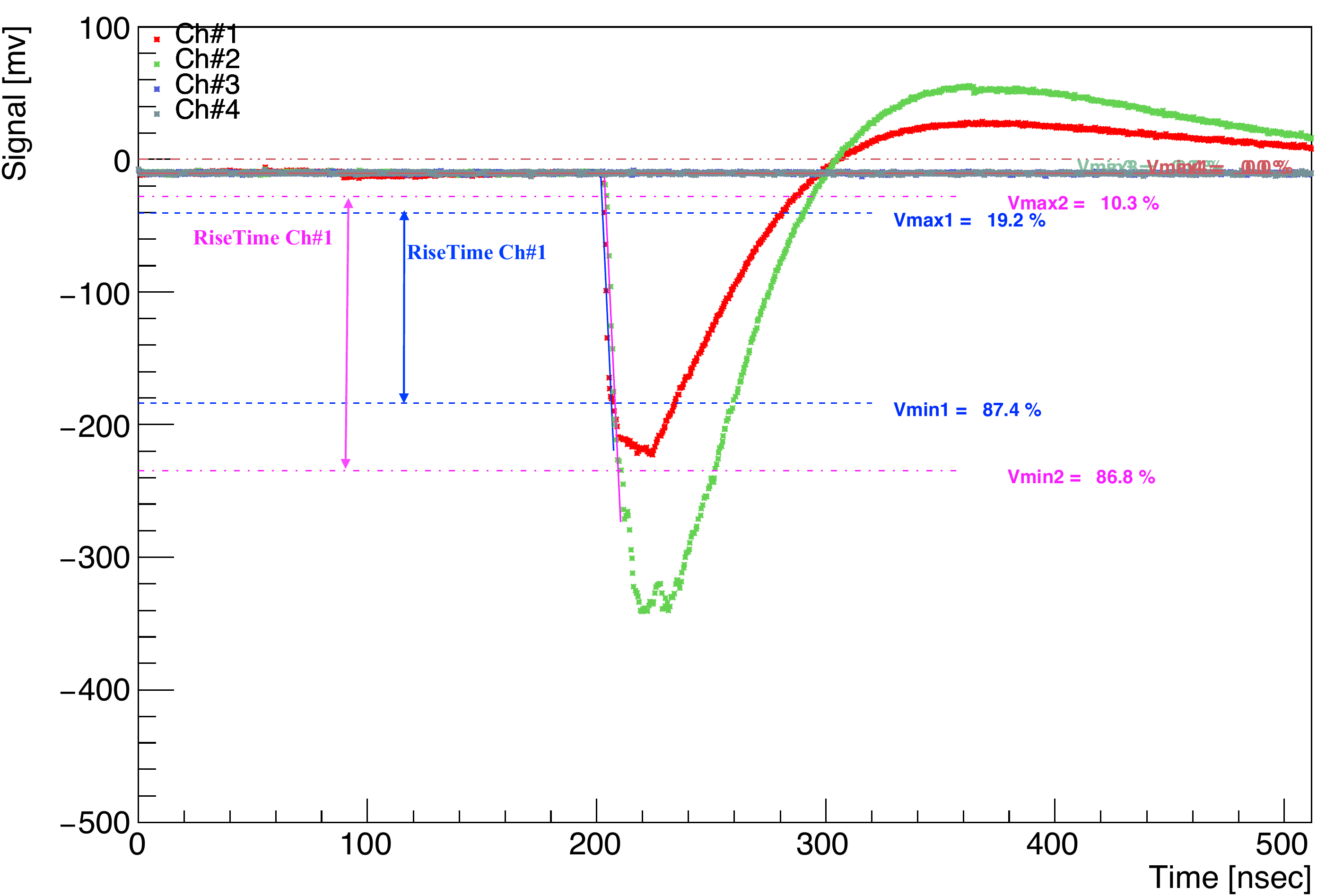}
  \caption{Signal shape of the rising time between the boundaries are $10\%$ percent of the amplitude and $90\%$ percent of the amplitude. } 
\label{fig:sipmSignal}
\end{figure}

\begin{figure}[hbt] 
 	\subfigure[Tile 1]{%
    		\includegraphics[width=0.48\textwidth]{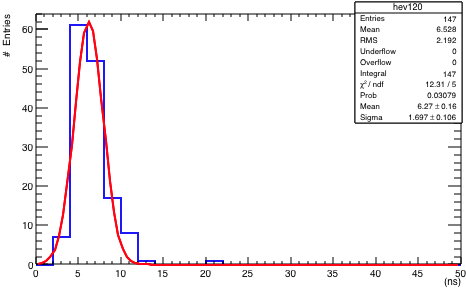}
    	} 
	\subfigure[Tile 2]{%
    		\includegraphics[width=0.48\textwidth]{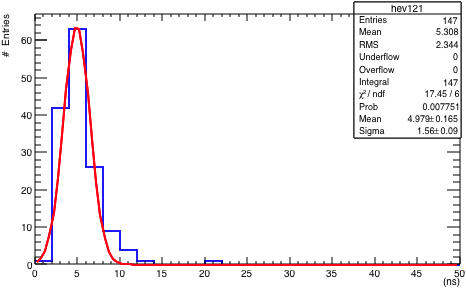}
    	}		
  	\caption{Rising time distribution of the signal is between $10\%$ and $90\%$ percent of the amplitude. (a) is for $Tile~1$ and in (b) is for $Tile~2$, respectively.} 
	\label{fig:risingTimeT}
\end{figure}

Time taken by the signal to rise from minimum level to maximum level is named the \textit{rise time}, and the time taken by the signal goes from maximum level to minimum level is named \textit{fall time}. The nonlinearity of the signal typically takes place at the bottom and at the top of the signal so that the rise time is generally determined between the $10\%$ and $90\%$ percent of the amplitude of the signal. A typical SiPM signal registered by the detector is seen in Figure~\ref{fig:sipmSignal}. It has a final rise and fall times.
The results are given in the Figure~\ref{fig:risingTimeT}.
\vspace{6cm}
\\In Figure~\ref{fig:risingTimeT}, the mean of the gaussian fit on the rising time distribution are $6.27 \pm 0.16~ns$ and $4.979\pm0.165~ns$ for $Tile~1$ and $Tile~2$, respectively. 

\subsubsection{Testing the robustness of the Fit }
\label{sec:robustnessFit}
\begin{figure}[!hbt] 
 	\subfigure[Tile 1]{%
    		\includegraphics[width=0.48\textwidth]{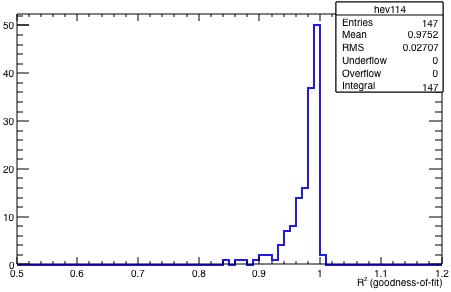}
    	}  
	\subfigure[Tile 2]{%
    		\includegraphics[width=0.48\textwidth]{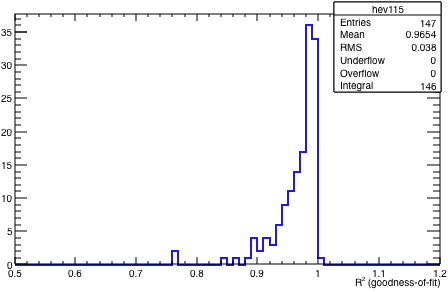}
    	}		
  	\caption{Goodness of Fit distribution of the signal is between $10\%$ and $90\%$ percent of the amplitude. (a) is for $Tile~1$ and in (b) is for $Tile~2$, respectively.} 
	\label{fig:rSquareT}
\end{figure}

Goodness of fit, $R^{2}$, is called correlation coefficient, is a quantity which indicates the quality of a least square fitting to the data. If $R^{2} = 1$ means the fit is perfect but generally expected that this is close to $1$.
Figure~\ref{fig:rSquareT} depicts that the $R^{2}$ distribution for all registered events for each tile. The mean value is $0.9752$ and $0.9654$ for $Tile~1$ and $Tile~2$, respectively. 

\subsection{Multiplicity of the Signal Test} \label{subsec:multiplicity}

\begin{figure}[hbt] 
 \centering
    \includegraphics[width=0.9\textwidth]{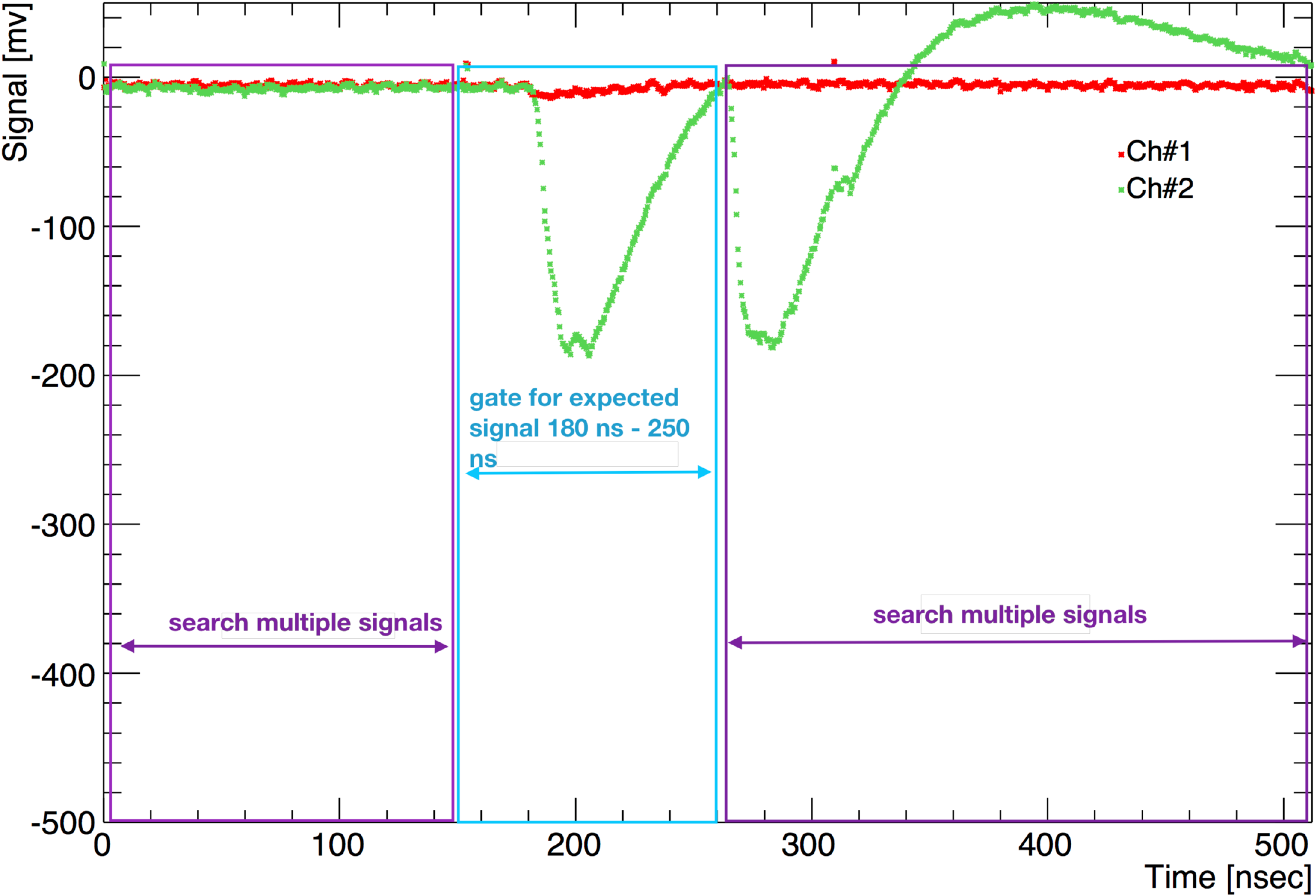}
  \caption{Multiplicity of registered SiPM signal in a full sampling window.}
\label{fig:multiplicityEvents}
\end{figure}

The signal (seen in Figure~\ref{fig:tof_method}) is a single cosmic ray (generally muon) points towards the region of the SiPMs. Cosmic rays are large collections of particles. Because these cosmic rays contain many particles, the particles will often separate, which could cause multiple hits in the region of the SiPM. One can see that there are two main hits in the given particular event (seen in Figure~\ref{fig:multiplicityEvents}).\\
\textit{Multiplicity}  is defined as a number of hits inside the full sampling window of the SiPMs had per event which is sketched in Figure~\ref{fig:multiplicityEvents}.
Since the detector is not limited to one characteristic of event, those both particles would be present, and point towards a nearby region, causing the total charge to be excessed.

\begin{figure}[hbt] 
 	\subfigure[]{%
    		\includegraphics[width=0.48\textwidth]{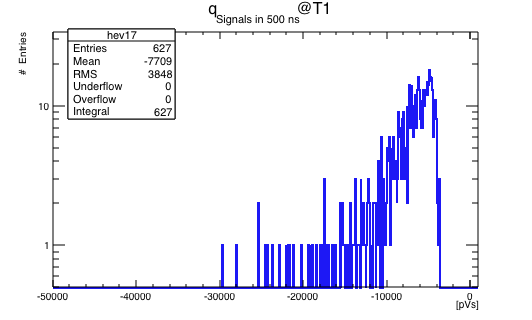}
    	}   
	\subfigure[]{%
    		\includegraphics[width=0.48\textwidth]{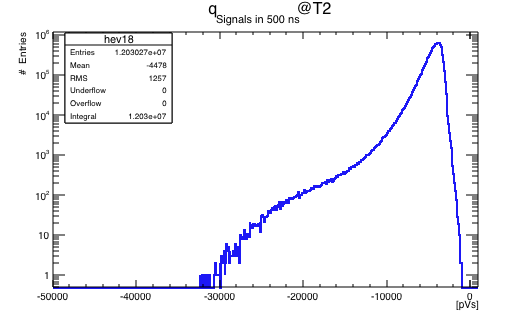}
    	}
	\subfigure[]{%
    		\includegraphics[width=0.48\textwidth]{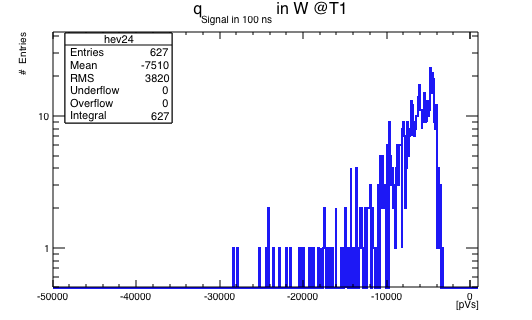}
    	}   
	\subfigure[]{%
    		\includegraphics[width=0.48\textwidth]{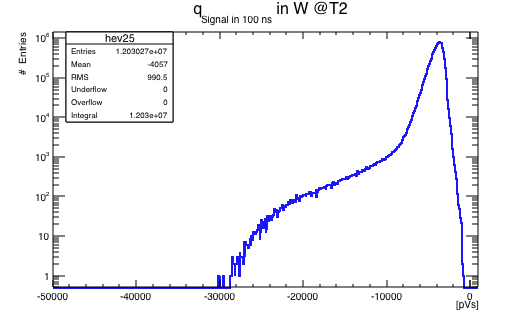}
    	}	
	\caption{Integrated charge for all signal where the baseline was removed from each signal. Histograms depicts that the excessive number of particles (a) is for $Tile~1$ and in (b) is for $Tile~2$, respectively. The integrated charge for gated signals (c) is for $Tile~1$ and in (d) is for $Tile~2$, respectively. DAQ trigger channel is $Tile~2$}.
	\label{fig:qSignals500nsT}
\end{figure}

In order to find the multiplicity of the signals,  the total charge of the expected signal in the window gate (180 ns - 250 ns) and the total charge of the all signal in the full sampling window (512 ns) were integrated and the baseline of the each signal was removed from the total charge.
Figure~\ref{fig:qSignals500nsT}  shows that the integrated charge for the signals in the full sampling window ((a) $Tile~1$ and (b) $Tile~2$) and the signals within the gated window ((c) $Tile~1$ and (d) $Tile~2$).

\begin{figure}[hbt] 
 	\subfigure[]{%
    		\includegraphics[width=0.48\textwidth]{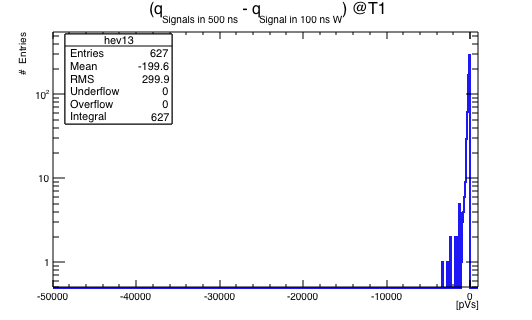}
    	}   
	\subfigure[]{%
    		\includegraphics[width=0.48\textwidth]{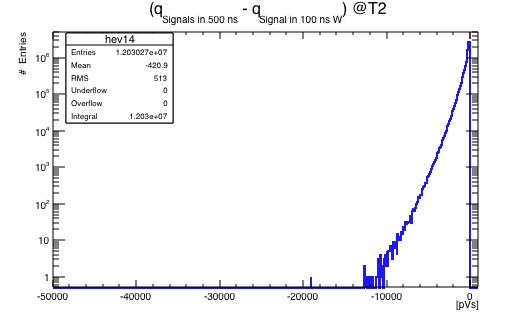}
    	}
	\subfigure[]{%
    		\includegraphics[width=0.48\textwidth]{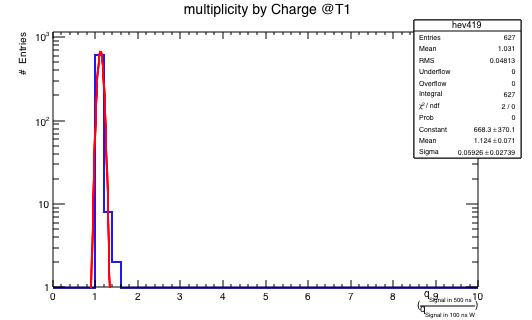}
    	}   
	\subfigure[]{%
    		\includegraphics[width=0.48\textwidth]{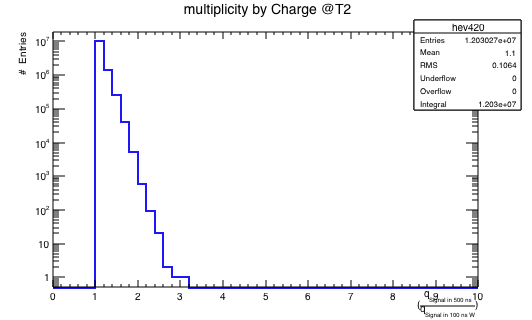}
    	}		
  	\caption{Multiplicity events in terms of integrated charge for the full sampling window (a) for $Tile~1$ and (b) for $Tile~2$,  and (c) and (d) give the multiplicity in terms of the number of events. DAQ trigger channel is $Tile~2$}.
	\label{fig:multiplicityByChargeDiffT}
\end{figure}

Figure~\ref{fig:multiplicityByChargeDiffT} ((a) Tile 1, (b) Tile 2), depicts that the multiplicity as a function of the integrated charge which was estimated by subtracting the total charge of the signal in the defined region from the total charge of the signals in the full sampling window. The multiplicity number as a function of the expected signal was also estimated and plotted in the Figure~\ref{fig:multiplicityByChargeDiffT}  ((c) Tile 1, (d) Tile 2).  Those values above 1 are due to more than one particle interacting with one cell. Since the $Tile~2$, was selected as a trigger in the DAQ, the number of registered signals were  statistically more than $Tile~1$ as shown in Figure~\ref{fig:qSignals500nsT} and Figure~\ref{fig:multiplicityByChargeDiffT}.

%%%%%%%%%%%%%%%%%%%%%%%%%%%%%%%%%%%%%%%%%%%%%%%%%%%%%%%%%%%%%%%%%%%
\section{Discussion and Conclusion} \label{sec:conclusion}
 In this paper, the performance study of the developed method for time-of-flight measurements using the scintillation counters are presented. A new DAQ system, described in section~\ref{subsec:DAQ}, working in automatically without needing a user for interaction. Moreover a new analysis program and a time-of-flight technique were developed and tested for further analysis. The rise time of the SiPM signals was accumulated and found to be $6.27\pm0.16$ and $4.979\pm0.165~ns$ for each tile. The coincidence timing resolution between the tiles were found about 1.7 ns signal from Figure~\ref{fig:TOF21}.  Multiplicity as a function of the integrated charge of the signals were estimated around  -119.6 pVs (-420 pVs) for $Tile~1$ ($Tile~2$). The limitation on the results is due to the present electronics used in the DAQ system. These results proves that the using this developed method satisfy very well the selection of direction of cosmic rays while removing the background.
 
% These results proves that the prototype detector is also capable of discriminating the particles going downward and upward direction using this developed method.

%%%%%%%%%%%%%%%%%%%%%%%%%%%%%%%%%%%%%%%%%%%%%%%%%%%%%%%%%%%%%%%%%%%
\section*{Acknowledgement} \label{sec:e}
 I would like to thank Prof. Dr. Haluk Denizli and Prof. Dr. Maurizio Iori for their support and encouragement during the course of this work.

%%%%%%%%%%%%%%%%%%%%%%%%%%%%%%%%%%%%%%%%%%%%%%%%%%%%%%%%%%%%%%%%%%%%%%%%
%\section*{References}
%\bibliographystyle{polonica} 
% unsrt 
\bibliographystyle{ieeetr}
\bibliography{mybibfile}

%uncomment the following lines to place a figure
%\begin{figure}[htb]
%\centerline{%
%\includegraphics[width=12.5cm]{Fig1}}
%\caption{Plot of ...}
%\label{Fig:F2H}
%\end{figure}

\end{document}